\documentclass[fleqn,10pt]{wlscirep}
\usepackage[utf8]{inputenc}
\usepackage[T1]{fontenc}
\title{Wideband Coherent Microwave Conversion via Magnon Nonlinearity in Hybrid Quantum System}

\author[1+]{Jiahao Wu}
\author[2+]{Jiacheng Liu}
\author[2+]{Zheyu Ren}
\author[1]{Man Yin Leung}
\author[1]{Wai Kuen Leung}
\author[1]{Kin On Ho}
\author[1]{Xiangrong Wang}
\author[1,2,3,*]{Qiming Shao}
\author[1,3,*]{Sen Yang}
\affil[1]{Department of Physics, The Hong Kong University of Science and Technology, Clear Water Bay, Hong Kong, China}
\affil[2]{Department of Electronic and Computer Engineering, The Hong Kong University of Science and Technology, Clear Water Bay, Hong Kong, China}
\affil[3]{Center for Complex Quantum Systems, The Hong Kong University of Science and Technology, Clear Water Bay, Hong Kong, China}

\affil[*]{eeqshao@ust.hk, phsyang@ust.hk}

\affil[+]{these authors contributed equally to this work}

\begin{abstract}
Frequency conversion is a widely realized physical process in nonlinear systems of optics and electronics. As an emerging nonlinear platform, spintronic devices have the potential to achieve stronger frequency conversion.
Here, we demonstrated a microwave frequency conversion method in a hybrid quantum system, integrating nitrogen-vacancy centers in diamond with magnetic thin film CoFeB. We achieve a conversion bandwidth ranging from 0.1 to 12GHz, presenting an up to $\mathrm{25^{th}}$ order frequency conversion and further display the application of this method for frequency detection and qubits coherent control. Distinct from traditional frequency conversion techniques based on nonlinear electric response, our approach employs nonlinear magnetic response in spintronic devices. 
The nonlinearity, originating from the symmetry breaking such as domain walls in magnetic films, presents that our method can be adapted to hybrid systems of other spintronic devices and spin qubits, expanding the application scope of spintronic devices and providing a promising on-chip platform for coupling quantum systems.

\end{abstract}
\begin{document}

\flushbottom
\maketitle

\thispagestyle{empty}

\section*{Introduction}

Frequency conversion between optical photons has been extensively studied in nonlinear optics, yielding significant applications, including the fabrication of lasers covering all spectra \cite{shen2003principles} and the coupling of multiple quantum systems \cite{Maring2017}.
Much research focuses on achieving conversion between microwave and optical photons to facilitate long-distance quantum communication \cite{Mirhosseini2020, Arnold2020}. 
For the coupling of different solid-state qubit systems, such as superconducting qubits and spin qubits, a typical requirement is frequency matching \cite{Scarlino2019, PhysRevX.8.041018}, which is not a common property for solid-state qubits\cite{RevModPhys.93.025005, RevModPhys.95.025003}. Solid-state qubits typically resonate with microwaves, which can range from several GHz to hundreds of GHz depending on the types\cite{PhysRevX.13.031022, Becker2016, THIERING20201}. 
These characteristics of solid-state qubits pose a challenge in coupling different quantum systems and building hybrid quantum networks\cite{Kimble2008}.
To couple solid-state qubits with different resonant frequencies, coherent conversion between microwave photons in the near field region is especially crucial.  
Moreover, wideband frequency conversion is also significant for quantum sensing, which can be understood as coupling qubits with the environment.
The main approach of microwave sensing involves measuring the spin relaxation time, $T_1$, of spin qubits\cite{doi:10.1126/science.aak9611, Hall2016, Casola2018, Ermakova2013}. 
This requires a large tunable bias field to shift the resonant frequency and align it with the detected signal \cite{https://doi.org/10.1002/adom.201600039, Magaletti2022}. Under the magnetic field of the Tesla scale, their resonance frequencies can shift by tens of GHz, attributed to the Zeeman effect\cite{10.1063/5.0055642}. Such intense tuning fields risk modifying the intrinsic dynamics of the samples under study, obscuring the phenomena of interest \cite{doi:10.1126/sciadv.abd3556, Thiel2016}. Moreover, the stringent aligned magnetic field greatly hinders the miniaturization of quantum sensing.
To solve these problems, we need a new method of microwave conversion that needs to be available over a wide bandwidth range to facilitate the coupling of different quantum systems. The conversion should be passive to minimize thermal noise from traditional active components. Additionally, it needs to be easily integrated on-chip with solid-state qubits to achieve the integration and miniaturization of hybrid quantum systems.

Here, we explore a frequency conversion method in the spintronic device that is capable of fulfilling the above requirements.
In the field of nonlinear optics, the nonlinear response of the electric field ($P=\chi^{\left (  1\right ) } E+\chi^{\left (  2\right ) } E^2+\cdots$) is often the subject of considerable interest\cite{Han:21}. The nonlinear electric response is usually weak in nonlinear optics crystals, manifesting as second-order nonlinearities\cite{shen2003principles}. In contrast, its counterpart, the nonlinear response of the magnetic field ($M=\chi_M^{\left (  1\right ) } H+\chi_M^{\left (  2\right ) } H^2+\cdots$), is relatively underexplored.
Some research indicates that strong interaction can easily occur between magnons in ferromagnetic media and microwave photons\cite{PhysRevB.93.144420}. 
Recently, Carmiggelt et al. observed a nonlinear four-wave mixing based on ferromagnetic resonance (FMR) in the YIG film\cite{Carmiggelt2023}. Koerner et al. discovered an up to 50th-order nonlinear harmonic signal, which is coming from switching effects in magnetic film, beyond the FMR region in NiFe thin film\cite{doi:10.1126/science.abm6044}. 
The presence of higher-order nonlinear magnetic response provides higher degrees of freedom in frequency conversion.  We propose a coherent frequency conversion method with much wider bandwidth by strong nonlinear response coming from symmetry breaking in domain walls rather than the magnons scattering in FMR region.
We demonstrate this method on a hybrid system integrating solid-state qubits with a spintronic device, taking nitrogen-vacancy (NV) center in diamond\cite{RevModPhys.92.015004} as an example of a solid-state qubit, and the $\mathrm{CoFeB}$ thin film on waveguide as an example of the spintronic device. First, the input microwaves generate corresponding magnons, through the linear magnetic dipole interaction. Then, the strong nonlinear response on the rich texture magnetic film results in the multi-wave mixing of magnons, and the converted magnons couple with NV centers.

We measure a wide-band microwave frequency conversion spectrum, covering a wide frequency range spanning from 100MHz to 12 GHz. The range is limited by our instrumentation, having the potential to reach tens of GHz. The spectrum shows that the spintronic microwave converter can achieve a flexible combination of two microwaves. Our experiments and simulations illustrate that the frequency conversion mechanism relies on nonlinearity,$\chi ^{\left (  2,3,\cdots \right ) }$, which originates from symmetry breaking in magnetic domain walls of the magnetic film. 
We display that our hybrid system can couple environmental signals with solid-state qubits, realizing wideband microwave sensing under a fixed magnetic field. This application dramatically enhances the quantum sensing bandwidth of solid-state qubits, constituting a major advance toward the precise characterization and miniaturization of microwave quantum sensing applications.
Furthermore, we achieve coherent quantum control of the solid-state spins by performing up-conversion. The pumping microwave photons are detuned from the electron spin resonance (ESR) frequency by a few GHz. This process reveals that the converted magnons retain good coherence. It shows that the frequency conversion in the hybrid system can be utilized to couple spin qubit systems. Subsequently, we obtain a competitive conversion efficiency (5.9\% for the third-order conversion) by analyzing Rabi frequencies. This solution, not only addresses the challenges in quantum information but also opens up a promising avenue for nonlinear spintronic devices.

\section*{Result}
\subsection*{Hybrid system}

\begin{figure}[htbp]
  \centering
  \includegraphics[width=\linewidth]{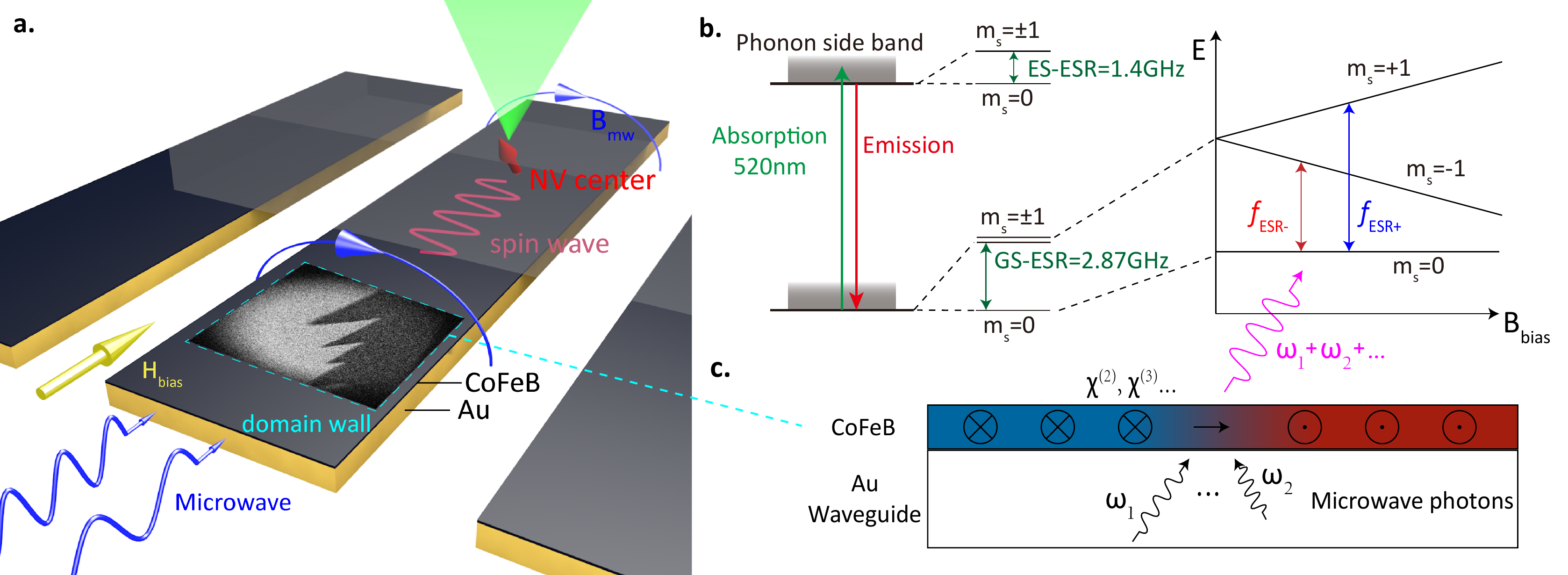}  
  \caption{\textbf{a.} The configuration of magnon-spin qubit hybrid system. The bulk diamond with implanted NV centers is closed to the magnetic film on the waveguide. The bias field is parallel to the waveguide. The two microwave signals show the delivery of the microwaves refer to Fig. \ref{fig:3} and \ref{fig:4}. \textbf{b.} The energy level of the NV center. Left panel: A 520nm green laser is used to initialize the spin states and the red photoluminescence (PL) is detected for the read-out of spin states. The denoted excited state electron spin resonance (ES-ESR) and ground state electron spin resonance (GS-ESR) correspond to the zero-field condition. Right panel: Under a bias field aligned with the NV-axis, the frequencies of GS-ESR will shift with the bias field due to the Zeeman effect. \textbf{c.} The conversion principle of the hybrid system. The microwave photons will stimulate the magnon (spin-wave) in the $\mathrm{CoFeB}$ film, and the presence of magnetic domain walls leads to strong nonlinear magnetic effects due to the nonlinear susceptibilities $\chi ^{\left (  2\right ) },\chi ^{\left (  3\right ) },\cdots $, generating multi-wave mixing.  Violet arrow line: The NV center can detect the stray fields generated by converted magnons when their frequency is resonant with the ESR frequencies $f_{ESR\pm}$ of NV centers.}
  \label{fig:1}
\end{figure}

Our hybrid system integrates NV centers with a 15nm $\mathrm{CoFeB}$ thin film deposited on a coplanar waveguide (CPW) in two configurations: nanodiamonds (ND) containing NV centers randomly dispersed on the CPW surface, and a bulk diamond with implanted nitrogen-vacancy centers placed close to CPW. Both coupling schemes yield similar results, illustrating the general applicability of the hybrid system for various solid-state spin systems.
The NV centers are situated close to the magnetic film, serving as a sensor of the stray fields generated by the spin waves (magnons). The spin waves in the magnetic thin film are excited by microwaves propagating through the gold waveguide beneath the film. Additionally, the applied static magnetic field is oriented along the direction of microwave propagation, resulting in a perpendicular alignment between the microwave field and the static magnetic field. The configuration of our hybrid system is depicted in figure \ref{fig:1}a.

\subsection*{Strong nonlinear effects and the accompanying frequency conversion in magnetic films }

\begin{figure}[htbp]
  \centering
  \includegraphics[width=\linewidth]{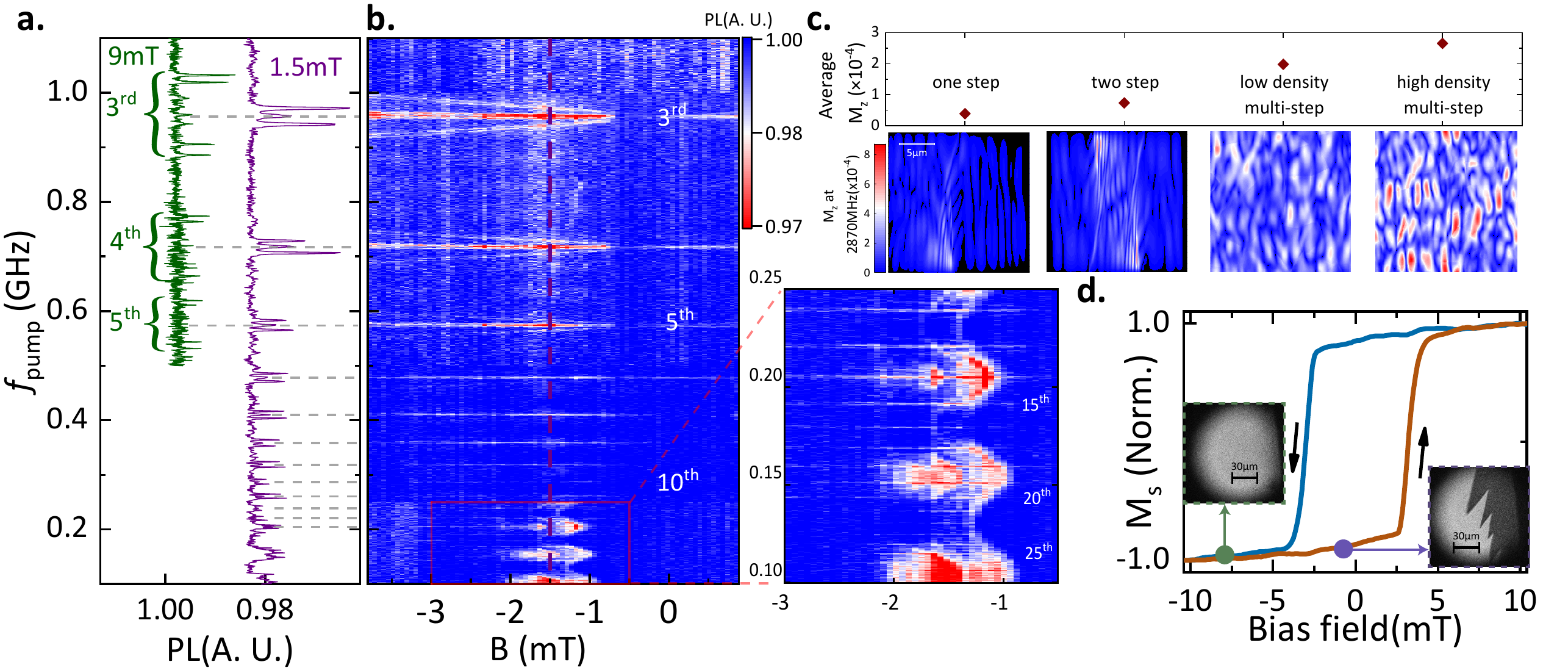}  
  \caption{\textbf{a.} ODMR spectrum at different bias field. The ODMR at 1.5 mT (solid purple line) corresponds to the cross-section along the dashed purple line in (b), where higher-order harmonic responses are observed under the bias field, indicating strong nonlinear effects. Dashed gray lines: The center of each harmonic peak, corresponding to integer fractions $1/n$ of ESR frequency $f_{ESR}$ = 2870 MHz. The ODMR at 9 mT (solid green line) represents the harmonic responses in the saturation magnetization region, indicating that the nonlinear response does not solely originate from domain walls. The depicted resonance peak splitting arises from the magnetic field-induced Zeeman splitting of the NV center's ESR frequency. \textbf{b.} Map of the ODMR signal vs driving microwave frequency $f_{pump}$ and magnetic field. Harmonics signals up to $25^{th}$ order are detected, with an enlarged image displayed on the right side.  \textbf{c.} Micromagnetic simulations of different domain configurations, where “one-step” is single domain wall, “two-step” is double domain wall, and “multi-step” is multi-domain (See Fig. S8 for details). The lower panels are the simulated spatial distribution of $M_z$ at 2870MHz of different domain configurations. All models are excited by microwave field at $f_{pump}$ = 287MHz. The upper panel shows the spatial average $M_z$ of each model. \textbf{d.} MOKE measurement. Hysteresis loop measured by in-plane MOKE measurement. The green arrow: MOKE image (inset) in the saturated magnetization region. The purple arrow: MOKE image (inset) in the abundant domain textures region.}
  \label{fig:2}
\end{figure}

Our first experiment focuses on detecting the high-order harmonic frequencies of microwaves induced by strong nonlinear responses in the CoFeB magnetic thin film, by optically detected magnetic resonance (ODMR)\cite{doi:10.1126/science.276.5321.2012}. 
The detection principle is that resonant microwave fields can drive ESR transitions ($f_{ESR}\sim$ 2.87 GHz @ 0mT) in the nitrogen-vacancy centers, leading to reductions in their photoluminescence (PL) intensity (see figure \ref{fig:1} b\&c). By sweeping the microwave frequencies, resonant microwaves elicit a clear decrease in PL  intensity, while non-resonant microwaves do not alter the PL intensity.
To analyze the condition facilitating magnetic responses within the ferromagnetic layer, we adjust the static bias field magnitude $B_{bias}$ and pump microwave frequency $f_{pump}$, resulting in the PL intensity being displayed as a function of the static bias field and pump microwave frequency, in figure \ref{fig:2} b. We initially saturate the magnetization of the magnetic thin film by applying a +10 mT field, then the field is adjusted to the negative direction. Normally, only microwaves at resonance conditions affect the PL intensity, and we should observe a set of slowly splitting peaks starting from $f_{ESR}$. However, in this experiment, we find that unlike usual, when the driving microwave's harmonic frequencies ($n\cdot f_{pump}$) align with the resonant conditions, a noticeable decrease in PL counts can also be observed. Specifically, it means that whenever we apply a driving microwave with a frequency of $f_{mw} = f_{ESR}/7 \sim $ 410MHz on the waveguide, the NV center can detect a resonant signal with a frequency of $f_{ESR} \sim $ 2870MHz, corresponding to the $\mathrm{7^{th}}$ harmonic of the driving microwave.
This implies that the magnetic texture in the spintronic device serves as a frequency multiplier, converting a non-resonant microwave signal to a harmonic signal resonant with the NV center.

All microwave components utilized in our experiment have undergone meticulous filtering procedures to ensure that the observed results stem solely from the nonlinear effects of the magnetic film, while eliminating any potential nonlinearities originating from the microwave devices. Given the NV center's role as a near-field signal sensor, we have conducted a comprehensive analysis by comparing not only the results obtained from coplanar waveguide (CPW) with and without a coated ferromagnetic (FM) thin film but also the NV centers positioned above the waveguide and in the gap region. These comparative experiments confirm that only the NV centers on CPW coated with an FM thin film can detect multiple harmonic frequencies, and this phenomenon of multiple harmonics originates from the CoFeB magnetic thin film. We observe at least $\mathrm{25^{th}}$ harmonic frequencies on CoFeB samples,  while previous work on soft magnetic NiFe thin films has observed up to $\mathrm{50^{th}}$ harmonic frequencies\cite{doi:10.1126/science.abm6044}. These imply that the spin wave harmonics effect appears to be a common feature of various ferromagnetic thin films.


The microwave photons emitted from the CPW stimulate dynamic magnetization oscillations in the interfaced ferromagnetic layer. 
From the perspective of nonlinear optics, the domain walls and the edges/interfaces of the magnetic film represent a symmetry break, resulting in a nonlinear susceptibility and a nonlinear magnetic response. 
In magnetic systems, the nonlinear magnetic response can be expressed in a universal formulation: 
\begin{align}
\label{eq:Mag}
\mathbf{M}(t) = \chi ^{\left (  1\right ) } \mathbf{H}(t)+ \chi ^{\left (  2\right ) } \mathbf{H}^2(t) + \chi ^{\left (  3\right ) } \mathbf{H}^3(t)+\cdots 
\end{align}
These nonlinear magnetic susceptibilities such as $\chi ^{\left (  2\right ) }$ and $\chi ^{\left (  3\right ) }$ come from the symmetry breaking in domain boundaries. The intensity of the nonlinear signal depends on the value of nonlinear coefficients.
We inferred that an increase in domain wall length leads to larger symmetry-breaking regions, enhancing the nonlinear signal.
A narrower domain wall will also correspondingly enhance nonlinear response.

We verified this hypothesis through micromagnetic simulation (See M4 in supplementary information). The lower panels Fig \ref{fig:2}c mainly show the spatial distribution of the longitudinal magnetization $M_z$ at $10^{th}$ harmonic (2870MHz) under the 287 MHz microwave excitation. Each sub-figure corresponds to a different static structure, where “one-step” is single domain wall, “two-step” is double domain wall, “low density multi-step” is lower density multi-domain, and “high density multi-step” is higher density multi-domain. Although there is a slight spatial non-uniformity in the $M_z$ across each model area, the spatial average $M_z$ of each model, as shown in the upper panel of Figure 2c, clearly indicates a positive correlation between the length of domain walls and the intensity of the harmonic response.  We also try to alter the width of the domain wall by adjusting parameters. We observed that a narrower domain wall corresponded to a stronger nonlinear effect (See Fig. S9 in supplementary information). The longer and narrower domain walls lead to bigger nonlinear terms $\chi ^{\left (  2\right ) }$, $\chi ^{\left (  3\right ) }$, $\cdots$. This observation aligns with our phenomenological theoretical analysis.

Experimental manipulation allows for the control of nonlinearities strength, we observe the magnetic field dependence in the harmonic orders of the microwave during the experiment, see Figure \ref{fig:2} b. 
We find that the nonlinear harmonic signals are most pronounced under the magnetic fields from -0.5 to -2.5 mT, revealing that the intensity of the nonlinear effects is related to the magnitude of the bias field. To figure out what's happening in this bias field range, we use the magneto-optic Kerr effect (MOKE) microscopy to map out the magnetic texture of the ferromagnetic thin film, finding that the domain walls are zigzag-shaped and abundant within this region. This result aligns with our simulated results. We give out a qualitative theory, which shows that the nonlinear responses decay with the order slowly (See M5 in Supplementary Information) in regions with abundant domain walls. The basic reason is that the spin waves have a small or no gap along domain walls, therefore the dynamic field can always resonate with domain walls\cite{PhysRevB.90.014414}.
Furthermore, nonlinear harmonics signals were observed in the saturated magnetization region (bias field = 9 mT), albeit with diminished intensity, as shown in Figure \ref{fig:2} a. At this point, the magnetic film is theoretically fully magnetized by the bias field, and the domain walls disappear accordingly. 
Our results demonstrate the existence of nonlinear sources beyond domain walls, but further experiments are needed to confirm the specific roles of interfaces and edges, to expand the operational range of nonlinear devices \cite{PhysRevB.70.094408}.

The observed magnetic nonlinear response, drawing parallels with the principles of nonlinear optics, is anticipated to generate the multi-wave mixing phenomena, extending beyond the production of mere harmonics \cite{PhysRevB.97.094421}. 
We consider that the microwave field exciting the second-order nonlinearity consists of two different frequency components, which we denote as:
\begin{align}
\label{eq:2MagField}
\mathbf{H}(t) =H_1 e^{-i\omega_1 t} + H_2 e^{-i\omega_2 t}+ c.c.
\end{align}
The second-order contribution to the nonlinear magnetization is:
\begin{align}
\label{eq:2OrderMag}
\mathbf{M}^{\left (  2\right ) }(t) &=  \chi ^{\left (  2\right ) } \mathbf{H}^2(t) \\
&=\chi ^{\left (  2\right ) } \left [  H_1^2e^{-2i\omega_1 t} + H_2^2e^{-2i\omega_2 t}+ 2 H_1 H_2 e^{-i\left ( \omega_1 + \omega_2\right ) t} +
2 H_1 H^*_2 e^{-i\left ( \omega_1 - \omega_2\right ) t} + c.c.
\right ] +2\chi ^{\left (  2\right ) }\left (  H_1 H_1^* + H_2 H_2^*\right )
\end{align}
The equation shows that $\chi ^{\left (  2\right )}$ contributes to three-wave mixing, while $\chi ^{\left (  3\right )}$ and other higher-order nonlinearities should contribute to more complex four-wave and multi-wave mixing \cite{shen2003principles}. 

The second experiment demonstrates our implementation of multi-wave mixing using two different microwave sources through our hybrid system.
We keep the bias field at 1.5mT, and two microwave sources, $f_{1}$ and $f_{2}$, are concurrently applied to the hybrid system. We try to sweep the frequencies of two sources. Only when the sum and difference frequencies of two source frequencies or their harmonic frequencies align with the ESR frequency, a remarkable decrease in PL intensity can be observed. 

\begin{figure}[htbp]
  \centering
  \includegraphics[width=\linewidth]{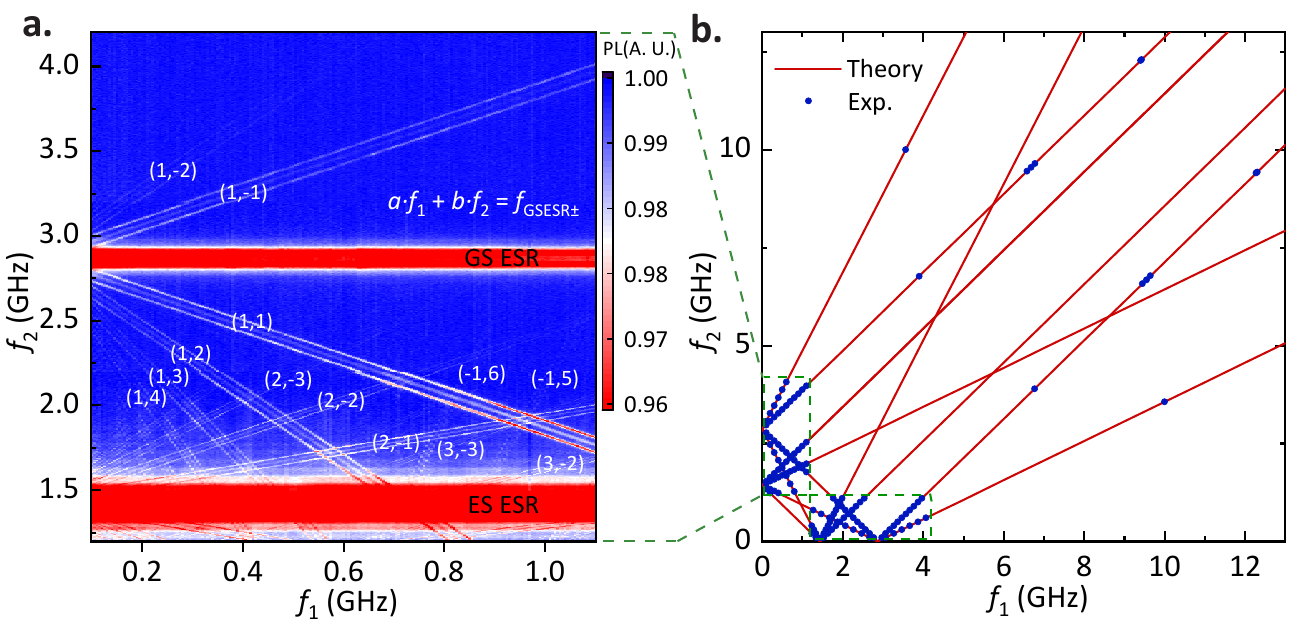}  
  \caption{\textbf{a.} ODMR measurement with two microwave sweeps under the 1.5mT magnetic field. Abundant multi-wave mixing is detected, the notations in the bracket stand for $(a,b)$ in equation (\ref{eq:Mixer}). The horizontal peaks marked correspond to the excited state ES-ESR transitions and the ground state GS-ESR transitions. \textbf{b.} Wideband spectrum of microwave mixing. Red line: Theoretical predicted signal of three-wave and four-wave mixing. Blue dot: Experiment detected signals. Green dashed box: Mark the experimental swept area in (\textbf{a}). }
  \label{fig:3}
\end{figure}

More precisely, the NV centers can detect the mixing signals whenever two microwave signals satisfy the following equations:
\begin{align}
\label{eq:Mixer}
a \cdot f_{1} + b \cdot f_{2} = f_{\pm ESR}
\end{align}
where $a$ and $b$ are integers.

In figure \ref{fig:3} a, we measure a 2-dimension spectrum of the non-linear spin wave response, the indicated notion means $(a,b)$ in equation \ref{eq:Mixer}. For this system, the two microwaves are interchangeable, hence the measurement points are symmetric about the line $f_{1}=f_{2}$. The measured spectrum demonstrates a rich set of frequency conversion paths, enabling a wide range of up- and down-conversion by flexibly combining harmonic generation and sum/difference frequency generation of different orders. 
In figure \ref{fig:3} b, we present the theoretical frequency spectrum of the hybrid system, and some specific experiment data points, demonstrating that our frequency conversion can span a range from 100MHz to 12GHz. The capabilities of our experimental instrumentation constrain the demonstrated frequency range. In
principle, a wider bandwidth can be achieved. We even observed $6^{th}$-order mixing ($3 \cdot f_{1} - 3 \cdot f_{2} = f_{ESR}$) between the third harmonic of $f_1$ and the third harmonic of $f_2$, which is rare in other nonlinear systems and highlights the advantage of frequency conversion in spintronic systems.

\subsection*{Wideband microwave sensing by frequency conversion}

Quantum sensing has demonstrated important applications such as nanoscale scanning magnetometers\cite{Balasubramanian2008, Maze2008, Maletinsky2012, doi:10.1126/science.1250113}, sensing under high pressure \cite{doi:10.1126/science.aaw4278,doi:10.1126/science.aaw4329, doi:10.1126/science.aaw4352}, and nanoscale nuclear magnetic resonance (NMR)\cite{mamin2013nanoscale,doi:10.1126/science.1231675, doi:10.1126/science.aal2538, doi:10.1126/science.aam8697}. Within the frequency range of 0-10 MHz, full range detection has been achieved through pulsed control methods, eliminating the necessity for tuning ESR frequencies\cite{RevModPhys.89.035002}. 
To measure weak alternating current (AC) signals higher than 10MHz, longitudinal spin relaxation time $T_1$ sensing is typically characterized, as it exhibits greater sensitivity to high-frequency signals. The conventional microwave sensing method involves tuning the bias magnetic field to alter the ESR frequency\cite{McCullian2020}. The signal can be detected only when the ESR frequency aligns with it, therefore the detection bandwidth depends on the tuning range of the bias field. We use a weak microwave to simulate the target signals coming from the environment, showing that our hybrid system can be used in varying the target frequency for T1 relaxometry, without the tunable magnetic field (see Figure. S2 in supplementary information).

\begin{figure}[htbp]
  \centering
  \includegraphics[width=0.8\linewidth]{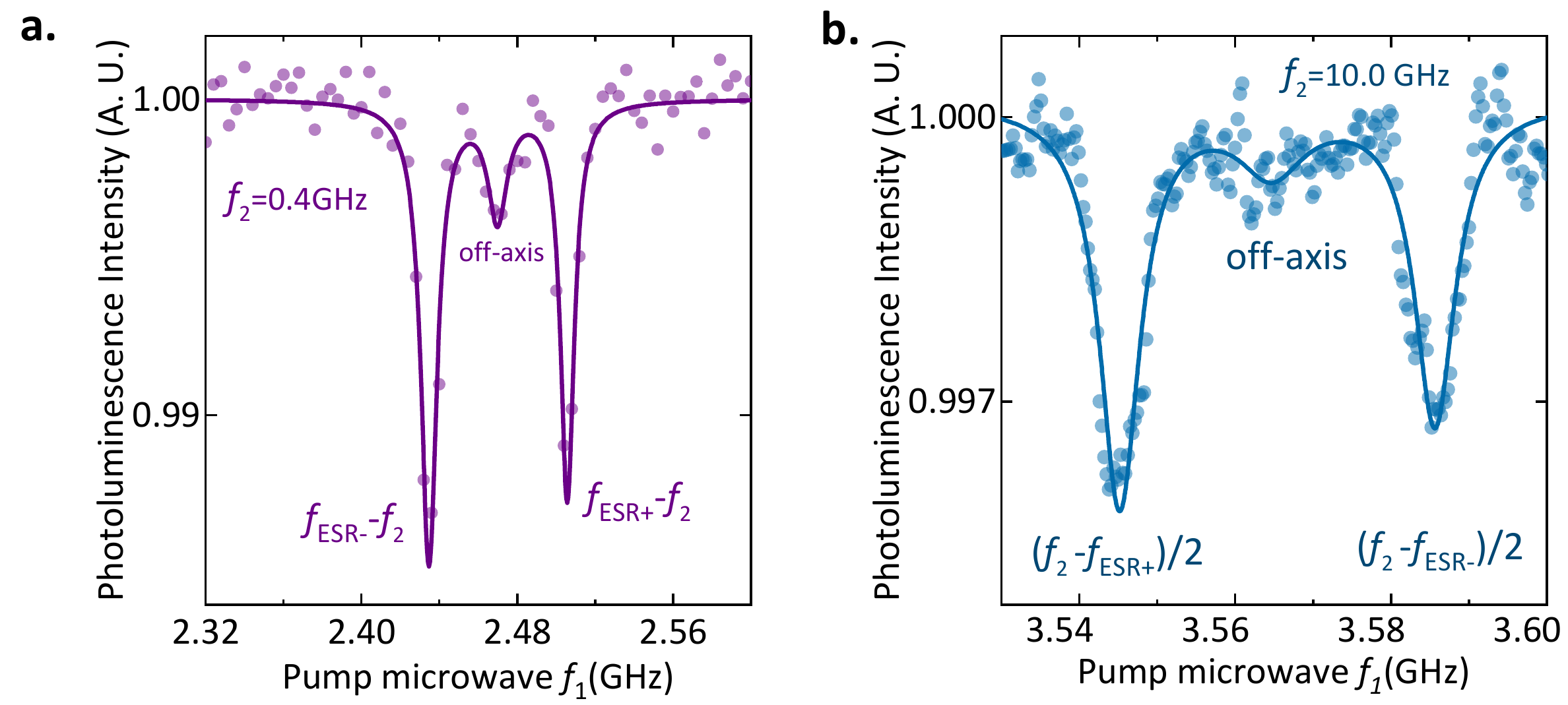}  
  \caption{\textbf{a.} Microwave sensing by up-conversion.   \textbf{b.} Microwave sensing by second-order down-conversion. The signal is detected by equation: $f_{2}=2 \cdot f_{1\mp}+f_{ESR\pm}$. 
  The off-axis peak shown in the figure is contributed by one of the four possible orientations of the NV axes, which is nearly perpendicular to the bias field. }
  \label{fig:4}
\end{figure}

We eliminate the need for complex externally applied bias magnetic fields through this hybrid system. By using the sum and difference frequencies generator, we try to convert the target signals into microwaves resonating with solid-state qubits, demonstrating a wideband microwave sensing under a fixed magnetic field. We perform the up-conversion microwave sensing in Figure \ref{fig:4} a.
We apply a continuous microwave $f_2 =$ 0.4 GHz to simulate an environment target signal. Then we apply a tunable pump microwave source $f_1$ to drive the nonlinear response of the magnetic device. With the magnetic sum frequency generator, we can detect the resonant peaks of the pump microwave, and derive the target signal using up-conversion protocol $f_1 + f_2 = f_{ESR}$ by NV center.

Another more noteworthy direction is the microwave sensing conducted by down-conversion. The detection of a high-frequency signal requires not only a strong tunable field but also a high-frequency pump source\cite{Carmiggelt2023}. Due to the multi-wave mixing properties of magnetic devices, we can achieve a variety of conversion protocols to further reduce the requirements for high-frequency signal detection. To detect a $f_2 =$ 10.0 GHz signal, through first-order down-conversion, we can use a scanning microwave source around $f_2-f_{ESR}=f_1=$ 7.13 GHz to read out the target signal. We can also realize second-order or higher-order down-conversion to compress the requirement of the pump source. Using the second-order down-conversion protocol $f_{2}=2 \cdot f_{1\mp}+f_{ESR\pm}$, we can now use more common, general-purpose microwave sources to accomplish the same task. It’s noteworthy that this approach also eliminates the need for a tunable 0.25T bias field, which is typically required in traditional detection methods. As our hybrid system can perform strong multi-wave mixing, every target signal $f_2$ can correspond to a spectrum fingerprint, which is related to the constant $F_{ESR}$ and resonant peaks $f_1\pm$ in pump frequency sweep measurement. 

So far, we have successfully illustrated a methodology that allows spin qubits to interact with other systems or environment signals by frequency conversion. The crux of this approach lies in the utilization of nonlinear magnons mixing for microwave frequency conversion. This, in turn, significantly broadens the detection spectrum of spin qubits, thereby enhancing the bandwidth of quantum sensing. What makes this method particularly intriguing is that the expansion of bandwidth is not reliant on an adjustable external magnetic field, which further enhances its applicability in various environments. The stabilization of the magnetic field ensures consistent and reliable sensing, thereby accurately characterizing the amplitude and frequency attributes of the detected microwave signal. The bandwidth of frequency conversion is contingent on the frequency of the magnons that can be excited in the material, with the upper limit potentially reaching tens of GHz.

\subsection*{Quantum coherent manipulation by frequency conversion}

To achieve coupling different solid-state qubit systems, we further substantiate that microwave photons, after frequency conversion, retain the same coherence as the source and are capable of executing quantum coherent control or coherently coupling quantum systems. This suggests that the microwave photons, derived from the conversion of nonlinear magnons, exhibit a high degree of coherence.

\begin{figure}[htbp]
  \centering
  \includegraphics[width=\linewidth]{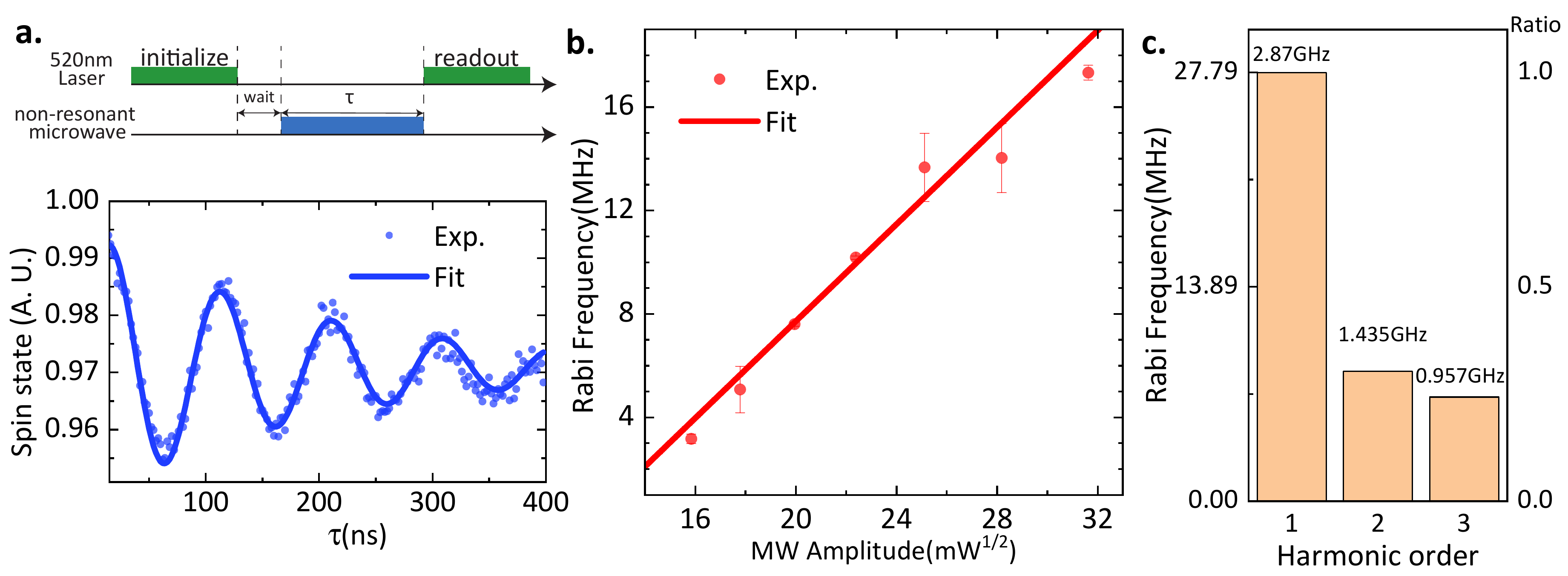}  
  \caption{\textbf{a.}Upper panel: Pulsed sequence of Rabi oscillation. Lower panel: Rabi oscillation driven by microwave at $f_{ESR}/3$ with the power of 27dBm.  \textbf{b.} Power dependence of Rabi frequencies driven by $\mathrm{3^{rd}}$ order harmonic, measured by NV centers in nanodiamonds. \textbf{c.} The Rabi oscillation frequency variation with different harmonic frequencies under the same output power.}
  \label{fig:5}
\end{figure}

Here we conduct a Rabi oscillation measurement by applying a microwave source at $f_{ESR}/3$, shown in figure \ref{fig:5}a. The Rabi oscillation frequency driven by the harmonic microwave exhibits a linear relationship with the amplitude of the driving microwave, see in figure \ref{fig:5}b. It indicates that within the power range, the amplitude of the driving microwave should also have a linear relationship with the amplitude of the converted harmonic signal. Our experimental results appear to deviate from our general estimation, as we expected the converted signal of the third-order harmonic to grow with the cube law of the amplitude of the driving microwave, as described by equation \ref{eq:Mag}. The current linear growth indicates that the conversion efficiency of our third-order harmonic conversion remains nearly constant. There are two possible reasons for this: first, a significant portion of the energy may have been converted into thermal magnons and dissipated; second, the system may be approaching a saturated conversion efficiency which is similar with the saturated effect in nonlinear optics. However, further analysis is required to determine the factors limiting this upper bound.

We further compare the Rabi oscillations driven by microwaves at $f_{ESR}$, $f_{ESR}/2$ and $f_{ESR}/3$ with identical power conditions, as depicted in \ref{fig:5}c. We can obtain the distribution of harmonic microwaves produced by the magnetic thin film after frequency conversion under a 20dBm microwave pump. While the remaining energy of the microwave at the pump frequency is regarded as 1, the energy ratio of the converted second-order harmonic microwave is 9.1\%, and that of the third-order harmonic microwave is 5.9\%. These results suggest that the frequency conversion efficiency gradually decreases with the increase in harmonic order, which is consistent with our simulation results. This Rabi frequency can precisely reflect the amplitude of the microwave field, providing a quantitative analysis tool for researching the nonlinear spin wave generation effect. 
The coherence is not only present in harmonic signals but also applicable to frequency mixing signals. By utilizing microwave sources with different frequencies, the frequency-mixing signals can also drive Rabi oscillations (see Fig. S1).

The realization of quantum coherent manipulation through frequency conversion demonstrates that the converted microwave exhibits strong phase preservation. Beyond enabling ultra-wideband signal detection, this hybrid system presents a viable pathway to enacting quantum coherent manipulations at non-resonant microwave frequencies, demonstrating the potential of coupling different solid-state qubit systems. For example, two long-lived quantum storage systems with different resonant frequencies can be connected to the hybrid system, which is critical for quantum network applications.

\section*{Discussion}
The nonlinear effects exhibited in the magnetic films we study are manifested in the strong high-order nonlinear susceptibility ($\chi ^{\left (  2\right ) }$, $\chi ^{\left (  3\right ) }$, $\cdots$) in equation \ref{eq:Mag}. These nonlinear coefficients are related to the symmetry of magnetism, such as the second-order nonlinearity resulting from the time-reversal symmetry breaking\cite{shen2003principles}. Compared to nonlinear crystals, our system can be very compact and integrated into various systems. This near-field frequency conversion also allows us to avoid the issue of phase matching, as the scale of our system is much smaller than the microwave wavelength.

We realize the coupling of spin qubits with the classical systems or environment, which is regarded as quantum sensing. To optimize the detection sensitivity of the solid-state quantum spin sensors, we can utilize weak microwave signal sensing protocols developed in the NV center, such as heterodyne detection scheme\cite{doi:10.1126/sciadv.abq8158, Meinel2021}, to improve the sensitivity. We can even convert the target signal to the range of 0-10MHz, and use more precise quantum sensing tools such as Ramsey interference\cite{Taylor2008} and dynamic decoupling\cite{PhysRevA.58.2733} to optimize measurement sensitivity. 
Furthermore, the microwave photons after frequency conversion exhibit surprisingly good coherence. This is sufficient for us to achieve quantum coherent control of various solid-state qubits using a non-resonant microwave source, and even realize the on-chip coupling between different quantum systems. For example, through our frequency conversion, the microwave photons emitted from the silicon vacancy($\mathrm{V_{Si}}$) in Silicon Carbide\cite{Kraus2014} can be transformed into microwave photons resonant with NV centers in diamond, thereby realizing the coupling between spin bits with different resonant frequencies.

Due to the fact that the nonlinear response is not solely coming from the symmetry breaking in domain walls, we can effectively engineer the nonlinear effects by meticulously modulating their magnetic texture and physical shapes to alter the nonlinear response in edges or interfaces of films\cite{PhysRevApplied.9.024029, Cenker2021}. 
For instance, based on simulation results (see Figure S8), we can modify the nonlinearity by changing the proportion of the material's magnetic parameters, such as the exchange stiffness coefficient $A_{ex}$ and uniaxial magnetic anisotropy $K_{u}$.
Furthermore, fabricating the devices with serrated edges and the utilization of thinner magnetic films, such as atomically thin van der Waals (vdW) magnetic materials \cite{Sun2021, Sun2019}, are expected to significantly enhance the nonlinear response of this hybrid system. Moreover, benefiting from the low power consumption and facile integration of spintronic devices, this hybrid system holds promise for integration with bio-sensing and cryogenic systems. These robust frequency conversions will significantly stimulate research on nonlinear spintronic devices, such as magnon IQ mixers and magnon frequency multipliers, opening up new avenues for the development of alternatives to traditional semiconductor devices.

\section*{Methods}
The $\mathrm{Au(100nm)/CoFeB(15nm)/TaO_x(2nm)}$ microscale coplanar waveguide (CPW), with $\mathrm{50 \mu m}$ width microstrip and $\mathrm{20 \mu m}$ width gap, was fabricated on a $\mathrm{SiO_2/Si}$ substrate utilizing magnetron sputtering technology, with $\mathrm{TaO_x}$ serving as a capping layer to inhibit oxidation. The CPW is connected to the circuit board through wire bonding, allowing for microwave propagation within the gold layer.

\subsection*{Micromagnetic simulation}

The simulations were performed using the Mumax3 software. The magnetic parameters adopted in our model align with the physical properties of realistic materials. During each simulation, the magnetization was stimulated by an external radio frequency (rf) field with a peak amplitude of $800 \mu T$ and a frequency of 287 MHz, oriented in the y-direction. Concurrently, a static bias field of 1.0 mT was applied in the x-direction. This configuration of the bias and rf fields was specifically chosen to replicate the orientation used in the actual experimental setup. To analyze the simulation results, a Fast Fourier Transform (FFT) was employed to extract the magnetization component $M_z$ at 2870 MHz from the simulated time evolution magnetization data. We then visualized the amplitude distribution of this component across the simulation area, as illustrated in Figure 2c.
Further details on the simulation parameters and settings can be found in the supplementary information.

\subsection*{Experimental setup}
In our study, we use NV centers in the bulk diamond and nanodiamonds as sensors. The diamond is implanted with 9.8 $\mathrm{keV}$ $\mathrm{^{15} N}$ ions at a dose of $\mathrm{2\cdot 10^{12} N/cm^2}$, resulting in the NV centers' concentration should be around 200ppm with a depth of around $\mathrm{10nm}$. The maximum distance between the NV centers and $\mathrm{CoFeB}$ is around $\mathrm{5\mu m}$, which is estimated from confocal imaging, constrained by particulate contamination (e.g. dust particles) at the interface between the diamond and $\mathrm{CoFeB}$ surfaces. Figure \ref{fig:2}, \ref{fig:3} and \ref{fig:4} are measured by NV centers in bulk diamond, and figure \ref{fig:5} is measured by NV centers in nanodiamonds. The detection of the stray field at ESR frequency was achieved by measuring the spin-dependent PL intensity under green laser excitation (520 nm) as depicted in Fig. \ref{fig:1}a. The PL signal was collected by a confocal microscopy system and detected by an avalanche photodetector (Excelitas SPCM-AQRH-10-FC). 
Due to the in-plane magnetic anisotropy (IMA) of magnetic film, the bias field is applied parallel to the thin film, and the current of the Helmholtz coil is controlled by a sourcemeter (Keithley 2400).

\section*{Data availability}
The data that support the findings of this work are provided within the main text and Supplementary Information. Data related to the work can also be made available from the corresponding author upon request. Preliminary results from this study were reported in the conference proceedings of the 2023 IEEE International Electron Devices Meeting (IEDM)\cite{10413860}.

\bibliography{sample}

\begin{thebibliography}{10}
\urlstyle{rm}
\expandafter\ifx\csname url\endcsname\relax
  \def\url#1{\texttt{#1}}\fi
\expandafter\ifx\csname urlprefix\endcsname\relax\def\urlprefix{URL }\fi
\expandafter\ifx\csname doiprefix\endcsname\relax\def\doiprefix{DOI: }\fi
\providecommand{\bibinfo}[2]{#2}
\providecommand{\eprint}[2][]{\url{#2}}

\bibitem{shen2003principles}
\bibinfo{author}{Shen, Y.}
\newblock \emph{\bibinfo{title}{The Principles of Nonlinear Optics}}.
\newblock Wiley classics library (\bibinfo{publisher}{Wiley}, \bibinfo{year}{2003}).

\bibitem{Maring2017}
\bibinfo{author}{Maring, N.} \emph{et~al.}
\newblock \bibinfo{journal}{\bibinfo{title}{Photonic quantum state transfer between a cold atomic gas and a crystal}}.
\newblock {\emph{\JournalTitle{Nature}}} \textbf{\bibinfo{volume}{551}}, \bibinfo{pages}{485--488}, \doiprefix\url{10.1038/nature24468} (\bibinfo{year}{2017}).

\bibitem{Mirhosseini2020}
\bibinfo{author}{Mirhosseini, M.}, \bibinfo{author}{Sipahigil, A.}, \bibinfo{author}{Kalaee, M.} \& \bibinfo{author}{Painter, O.}
\newblock \bibinfo{journal}{\bibinfo{title}{Superconducting qubit to optical photon transduction}}.
\newblock {\emph{\JournalTitle{Nature}}} \textbf{\bibinfo{volume}{588}}, \bibinfo{pages}{599--603}, \doiprefix\url{10.1038/s41586-020-3038-6} (\bibinfo{year}{2020}).

\bibitem{Arnold2020}
\bibinfo{author}{Arnold, G.} \emph{et~al.}
\newblock \bibinfo{journal}{\bibinfo{title}{Converting microwave and telecom photons with a silicon photonic nanomechanical interface}}.
\newblock {\emph{\JournalTitle{Nature Communications}}} \textbf{\bibinfo{volume}{11}}, \bibinfo{pages}{4460}, \doiprefix\url{10.1038/s41467-020-18269-z} (\bibinfo{year}{2020}).

\bibitem{Scarlino2019}
\bibinfo{author}{Scarlino, P.} \emph{et~al.}
\newblock \bibinfo{journal}{\bibinfo{title}{Coherent microwave-photon-mediated coupling between a semiconductor and a superconducting qubit}}.
\newblock {\emph{\JournalTitle{Nature Communications}}} \textbf{\bibinfo{volume}{10}}, \bibinfo{pages}{3011}, \doiprefix\url{10.1038/s41467-019-10798-6} (\bibinfo{year}{2019}).

\bibitem{PhysRevX.8.041018}
\bibinfo{author}{van Woerkom, D.~J.} \emph{et~al.}
\newblock \bibinfo{journal}{\bibinfo{title}{Microwave photon-mediated interactions between semiconductor qubits}}.
\newblock {\emph{\JournalTitle{Phys. Rev. X}}} \textbf{\bibinfo{volume}{8}}, \bibinfo{pages}{041018}, \doiprefix\url{10.1103/PhysRevX.8.041018} (\bibinfo{year}{2018}).

\bibitem{RevModPhys.93.025005}
\bibinfo{author}{Blais, A.}, \bibinfo{author}{Grimsmo, A.~L.}, \bibinfo{author}{Girvin, S.~M.} \& \bibinfo{author}{Wallraff, A.}
\newblock \bibinfo{journal}{\bibinfo{title}{Circuit quantum electrodynamics}}.
\newblock {\emph{\JournalTitle{Rev. Mod. Phys.}}} \textbf{\bibinfo{volume}{93}}, \bibinfo{pages}{025005}, \doiprefix\url{10.1103/RevModPhys.93.025005} (\bibinfo{year}{2021}).

\bibitem{RevModPhys.95.025003}
\bibinfo{author}{Burkard, G.}, \bibinfo{author}{Ladd, T.~D.}, \bibinfo{author}{Pan, A.}, \bibinfo{author}{Nichol, J.~M.} \& \bibinfo{author}{Petta, J.~R.}
\newblock \bibinfo{journal}{\bibinfo{title}{Semiconductor spin qubits}}.
\newblock {\emph{\JournalTitle{Rev. Mod. Phys.}}} \textbf{\bibinfo{volume}{95}}, \bibinfo{pages}{025003}, \doiprefix\url{10.1103/RevModPhys.95.025003} (\bibinfo{year}{2023}).

\bibitem{PhysRevX.13.031022}
\bibinfo{author}{Rosenthal, E.~I.} \emph{et~al.}
\newblock \bibinfo{journal}{\bibinfo{title}{Microwave spin control of a tin-vacancy qubit in diamond}}.
\newblock {\emph{\JournalTitle{Phys. Rev. X}}} \textbf{\bibinfo{volume}{13}}, \bibinfo{pages}{031022}, \doiprefix\url{10.1103/PhysRevX.13.031022} (\bibinfo{year}{2023}).

\bibitem{Becker2016}
\bibinfo{author}{Becker, J.~N.}, \bibinfo{author}{G{\"o}rlitz, J.}, \bibinfo{author}{Arend, C.}, \bibinfo{author}{Markham, M.} \& \bibinfo{author}{Becher, C.}
\newblock \bibinfo{journal}{\bibinfo{title}{Ultrafast all-optical coherent control of single silicon vacancy colour centres in diamond}}.
\newblock {\emph{\JournalTitle{Nature Communications}}} \textbf{\bibinfo{volume}{7}}, \bibinfo{pages}{13512}, \doiprefix\url{10.1038/ncomms13512} (\bibinfo{year}{2016}).

\bibitem{THIERING20201}
\bibinfo{author}{Thiering, G.} \& \bibinfo{author}{Gali, A.}
\newblock \bibinfo{title}{Chapter one - color centers in diamond for quantum applications}.
\newblock In \bibinfo{editor}{Nebel, C.~E.}, \bibinfo{editor}{Aharonovich, I.}, \bibinfo{editor}{Mizuochi, N.} \& \bibinfo{editor}{Hatano, M.} (eds.) \emph{\bibinfo{booktitle}{Diamond for Quantum Applications Part 1}}, vol. \bibinfo{volume}{103} of \emph{\bibinfo{series}{Semiconductors and Semimetals}}, \bibinfo{pages}{1--36}, \doiprefix\url{https://doi.org/10.1016/bs.semsem.2020.03.001} (\bibinfo{publisher}{Elsevier}, \bibinfo{year}{2020}).

\bibitem{Kimble2008}
\bibinfo{author}{Kimble, H.~J.}
\newblock \bibinfo{journal}{\bibinfo{title}{The quantum internet}}.
\newblock {\emph{\JournalTitle{Nature}}} \textbf{\bibinfo{volume}{453}}, \bibinfo{pages}{1023--1030}, \doiprefix\url{10.1038/nature07127} (\bibinfo{year}{2008}).

\bibitem{doi:10.1126/science.aak9611}
\bibinfo{author}{Du, C.} \emph{et~al.}
\newblock \bibinfo{journal}{\bibinfo{title}{Control and local measurement of the spin chemical potential in a magnetic insulator}}.
\newblock {\emph{\JournalTitle{Science}}} \textbf{\bibinfo{volume}{357}}, \bibinfo{pages}{195--198}, \doiprefix\url{10.1126/science.aak9611} (\bibinfo{year}{2017}).
\newblock \eprint{https://www.science.org/doi/pdf/10.1126/science.aak9611}.

\bibitem{Hall2016}
\bibinfo{author}{Hall, L.~T.} \emph{et~al.}
\newblock \bibinfo{journal}{\bibinfo{title}{Detection of nanoscale electron spin resonance spectra demonstrated using nitrogen-vacancy centre probes in diamond}}.
\newblock {\emph{\JournalTitle{Nature Communications}}} \textbf{\bibinfo{volume}{7}}, \bibinfo{pages}{10211}, \doiprefix\url{10.1038/ncomms10211} (\bibinfo{year}{2016}).

\bibitem{Casola2018}
\bibinfo{author}{Casola, F.}, \bibinfo{author}{van~der Sar, T.} \& \bibinfo{author}{Yacoby, A.}
\newblock \bibinfo{journal}{\bibinfo{title}{Probing condensed matter physics with magnetometry based on nitrogen-vacancy centres in diamond}}.
\newblock {\emph{\JournalTitle{Nature Reviews Materials}}} \textbf{\bibinfo{volume}{3}}, \bibinfo{pages}{17088}, \doiprefix\url{10.1038/natrevmats.2017.88} (\bibinfo{year}{2018}).

\bibitem{Ermakova2013}
\bibinfo{author}{Ermakova, A.} \emph{et~al.}
\newblock \bibinfo{journal}{\bibinfo{title}{Detection of a few metallo-protein molecules using color centers in nanodiamonds}}.
\newblock {\emph{\JournalTitle{Nano Letters}}} \textbf{\bibinfo{volume}{13}}, \bibinfo{pages}{3305--3309}, \doiprefix\url{10.1021/nl4015233} (\bibinfo{year}{2013}).

\bibitem{https://doi.org/10.1002/adom.201600039}
\bibinfo{author}{Shao, L.} \emph{et~al.}
\newblock \bibinfo{journal}{\bibinfo{title}{Wide-field optical microscopy of microwave fields using nitrogen-vacancy centers in diamonds}}.
\newblock {\emph{\JournalTitle{Advanced Optical Materials}}} \textbf{\bibinfo{volume}{4}}, \bibinfo{pages}{1075--1080}, \doiprefix\url{https://doi.org/10.1002/adom.201600039} (\bibinfo{year}{2016}).
\newblock \eprint{https://onlinelibrary.wiley.com/doi/pdf/10.1002/adom.201600039}.

\bibitem{Magaletti2022}
\bibinfo{author}{Magaletti, S.}, \bibinfo{author}{Mayer, L.}, \bibinfo{author}{Roch, J.-F.} \& \bibinfo{author}{Debuisschert, T.}
\newblock \bibinfo{journal}{\bibinfo{title}{A quantum radio frequency signal analyzer based on nitrogen vacancy centers in diamond}}.
\newblock {\emph{\JournalTitle{Communications Engineering}}} \textbf{\bibinfo{volume}{1}}, \bibinfo{pages}{19}, \doiprefix\url{10.1038/s44172-022-00017-4} (\bibinfo{year}{2022}).

\bibitem{10.1063/5.0055642}
\bibinfo{author}{Fortman, B.} \emph{et~al.}
\newblock \bibinfo{journal}{\bibinfo{title}{{Electron–electron double resonance detected NMR spectroscopy using ensemble NV centers at 230GHz and 8.3T}}}.
\newblock {\emph{\JournalTitle{Journal of Applied Physics}}} \textbf{\bibinfo{volume}{130}}, \bibinfo{pages}{083901}, \doiprefix\url{10.1063/5.0055642} (\bibinfo{year}{2021}).
\newblock \eprint{https://pubs.aip.org/aip/jap/article-pdf/doi/10.1063/5.0055642/15267976/083901\_1\_online.pdf}.

\bibitem{doi:10.1126/sciadv.abd3556}
\bibinfo{author}{Bertelli, I.} \emph{et~al.}
\newblock \bibinfo{journal}{\bibinfo{title}{Magnetic resonance imaging of spin-wave transport and interference in a magnetic insulator}}.
\newblock {\emph{\JournalTitle{Science Advances}}} \textbf{\bibinfo{volume}{6}}, \bibinfo{pages}{eabd3556}, \doiprefix\url{10.1126/sciadv.abd3556} (\bibinfo{year}{2020}).
\newblock \eprint{https://www.science.org/doi/pdf/10.1126/sciadv.abd3556}.

\bibitem{Thiel2016}
\bibinfo{author}{Thiel, L.} \emph{et~al.}
\newblock \bibinfo{journal}{\bibinfo{title}{Quantitative nanoscale vortex imaging using a cryogenic quantum magnetometer}}.
\newblock {\emph{\JournalTitle{Nature Nanotechnology}}} \textbf{\bibinfo{volume}{11}}, \bibinfo{pages}{677--681}, \doiprefix\url{10.1038/nnano.2016.63} (\bibinfo{year}{2016}).

\bibitem{Han:21}
\bibinfo{author}{Han, X.}, \bibinfo{author}{Fu, W.}, \bibinfo{author}{Zou, C.-L.}, \bibinfo{author}{Jiang, L.} \& \bibinfo{author}{Tang, H.~X.}
\newblock \bibinfo{journal}{\bibinfo{title}{Microwave-optical quantum frequency conversion}}.
\newblock {\emph{\JournalTitle{Optica}}} \textbf{\bibinfo{volume}{8}}, \bibinfo{pages}{1050--1064}, \doiprefix\url{10.1364/OPTICA.425414} (\bibinfo{year}{2021}).

\bibitem{PhysRevB.93.144420}
\bibinfo{author}{Bourhill, J.}, \bibinfo{author}{Kostylev, N.}, \bibinfo{author}{Goryachev, M.}, \bibinfo{author}{Creedon, D.~L.} \& \bibinfo{author}{Tobar, M.~E.}
\newblock \bibinfo{journal}{\bibinfo{title}{Ultrahigh cooperativity interactions between magnons and resonant photons in a yig sphere}}.
\newblock {\emph{\JournalTitle{Phys. Rev. B}}} \textbf{\bibinfo{volume}{93}}, \bibinfo{pages}{144420}, \doiprefix\url{10.1103/PhysRevB.93.144420} (\bibinfo{year}{2016}).

\bibitem{Carmiggelt2023}
\bibinfo{author}{Carmiggelt, J.~J.} \emph{et~al.}
\newblock \bibinfo{journal}{\bibinfo{title}{Broadband microwave detection using electron spins in a hybrid diamond-magnet sensor chip}}.
\newblock {\emph{\JournalTitle{Nature Communications}}} \textbf{\bibinfo{volume}{14}}, \bibinfo{pages}{490}, \doiprefix\url{10.1038/s41467-023-36146-3} (\bibinfo{year}{2023}).

\bibitem{doi:10.1126/science.abm6044}
\bibinfo{author}{Koerner, C.} \emph{et~al.}
\newblock \bibinfo{journal}{\bibinfo{title}{Frequency multiplication by collective nanoscale spin-wave dynamics}}.
\newblock {\emph{\JournalTitle{Science}}} \textbf{\bibinfo{volume}{375}}, \bibinfo{pages}{1165--1169}, \doiprefix\url{10.1126/science.abm6044} (\bibinfo{year}{2022}).
\newblock \eprint{https://www.science.org/doi/pdf/10.1126/science.abm6044}.

\bibitem{RevModPhys.92.015004}
\bibinfo{author}{Barry, J.~F.} \emph{et~al.}
\newblock \bibinfo{journal}{\bibinfo{title}{Sensitivity optimization for nv-diamond magnetometry}}.
\newblock {\emph{\JournalTitle{Rev. Mod. Phys.}}} \textbf{\bibinfo{volume}{92}}, \bibinfo{pages}{015004}, \doiprefix\url{10.1103/RevModPhys.92.015004} (\bibinfo{year}{2020}).

\bibitem{doi:10.1126/science.276.5321.2012}
\bibinfo{author}{Gruber, A.} \emph{et~al.}
\newblock \bibinfo{journal}{\bibinfo{title}{Scanning confocal optical microscopy and magnetic resonance on single defect centers}}.
\newblock {\emph{\JournalTitle{Science}}} \textbf{\bibinfo{volume}{276}}, \bibinfo{pages}{2012--2014}, \doiprefix\url{10.1126/science.276.5321.2012} (\bibinfo{year}{1997}).
\newblock \eprint{https://www.science.org/doi/pdf/10.1126/science.276.5321.2012}.

\bibitem{PhysRevB.90.014414}
\bibinfo{author}{Wang, X.~S.} \& \bibinfo{author}{Wang, X.~R.}
\newblock \bibinfo{journal}{\bibinfo{title}{Thermodynamic theory for thermal-gradient-driven domain-wall motion}}.
\newblock {\emph{\JournalTitle{Phys. Rev. B}}} \textbf{\bibinfo{volume}{90}}, \bibinfo{pages}{014414}, \doiprefix\url{10.1103/PhysRevB.90.014414} (\bibinfo{year}{2014}).

\bibitem{PhysRevB.70.094408}
\bibinfo{author}{Hansteen, F.}, \bibinfo{author}{Hunderi, O.}, \bibinfo{author}{Johansen, T.~H.}, \bibinfo{author}{Kirilyuk, A.} \& \bibinfo{author}{Rasing, T.}
\newblock \bibinfo{journal}{\bibinfo{title}{Selective surface/interface characterization of thin garnet films by magnetization-induced second-harmonic generation}}.
\newblock {\emph{\JournalTitle{Phys. Rev. B}}} \textbf{\bibinfo{volume}{70}}, \bibinfo{pages}{094408}, \doiprefix\url{10.1103/PhysRevB.70.094408} (\bibinfo{year}{2004}).

\bibitem{PhysRevB.97.094421}
\bibinfo{author}{Zhang, B.}, \bibinfo{author}{Wang, Z.}, \bibinfo{author}{Cao, Y.}, \bibinfo{author}{Yan, P.} \& \bibinfo{author}{Wang, X.~R.}
\newblock \bibinfo{journal}{\bibinfo{title}{Eavesdropping on spin waves inside the domain-wall nanochannel via three-magnon processes}}.
\newblock {\emph{\JournalTitle{Phys. Rev. B}}} \textbf{\bibinfo{volume}{97}}, \bibinfo{pages}{094421}, \doiprefix\url{10.1103/PhysRevB.97.094421} (\bibinfo{year}{2018}).

\bibitem{Balasubramanian2008}
\bibinfo{author}{Balasubramanian, G.} \emph{et~al.}
\newblock \bibinfo{journal}{\bibinfo{title}{Nanoscale imaging magnetometry with diamond spins under ambient conditions}}.
\newblock {\emph{\JournalTitle{Nature}}} \textbf{\bibinfo{volume}{455}}, \bibinfo{pages}{648--651}, \doiprefix\url{10.1038/nature07278} (\bibinfo{year}{2008}).

\bibitem{Maze2008}
\bibinfo{author}{Maze, J.~R.} \emph{et~al.}
\newblock \bibinfo{journal}{\bibinfo{title}{Nanoscale magnetic sensing with an individual electronic spin in diamond}}.
\newblock {\emph{\JournalTitle{Nature}}} \textbf{\bibinfo{volume}{455}}, \bibinfo{pages}{644--647}, \doiprefix\url{10.1038/nature07279} (\bibinfo{year}{2008}).

\bibitem{Maletinsky2012}
\bibinfo{author}{Maletinsky, P.} \emph{et~al.}
\newblock \bibinfo{journal}{\bibinfo{title}{A robust scanning diamond sensor for nanoscale imaging with single nitrogen-vacancy centres}}.
\newblock {\emph{\JournalTitle{Nature Nanotechnology}}} \textbf{\bibinfo{volume}{7}}, \bibinfo{pages}{320--324}, \doiprefix\url{10.1038/nnano.2012.50} (\bibinfo{year}{2012}).

\bibitem{doi:10.1126/science.1250113}
\bibinfo{author}{Tetienne, J.-P.} \emph{et~al.}
\newblock \bibinfo{journal}{\bibinfo{title}{Nanoscale imaging and control of domain-wall hopping with a nitrogen-vacancy center microscope}}.
\newblock {\emph{\JournalTitle{Science}}} \textbf{\bibinfo{volume}{344}}, \bibinfo{pages}{1366--1369}, \doiprefix\url{10.1126/science.1250113} (\bibinfo{year}{2014}).
\newblock \eprint{https://www.science.org/doi/pdf/10.1126/science.1250113}.

\bibitem{doi:10.1126/science.aaw4278}
\bibinfo{author}{Yip, K.~Y.} \emph{et~al.}
\newblock \bibinfo{journal}{\bibinfo{title}{Measuring magnetic field texture in correlated electron systems under extreme conditions}}.
\newblock {\emph{\JournalTitle{Science}}} \textbf{\bibinfo{volume}{366}}, \bibinfo{pages}{1355--1359}, \doiprefix\url{10.1126/science.aaw4278} (\bibinfo{year}{2019}).
\newblock \eprint{https://www.science.org/doi/pdf/10.1126/science.aaw4278}.

\bibitem{doi:10.1126/science.aaw4329}
\bibinfo{author}{Lesik, M.} \emph{et~al.}
\newblock \bibinfo{journal}{\bibinfo{title}{Magnetic measurements on micrometer-sized samples under high pressure using designed nv centers}}.
\newblock {\emph{\JournalTitle{Science}}} \textbf{\bibinfo{volume}{366}}, \bibinfo{pages}{1359--1362}, \doiprefix\url{10.1126/science.aaw4329} (\bibinfo{year}{2019}).
\newblock \eprint{https://www.science.org/doi/pdf/10.1126/science.aaw4329}.

\bibitem{doi:10.1126/science.aaw4352}
\bibinfo{author}{Hsieh, S.} \emph{et~al.}
\newblock \bibinfo{journal}{\bibinfo{title}{Imaging stress and magnetism at high pressures using a nanoscale quantum sensor}}.
\newblock {\emph{\JournalTitle{Science}}} \textbf{\bibinfo{volume}{366}}, \bibinfo{pages}{1349--1354}, \doiprefix\url{10.1126/science.aaw4352} (\bibinfo{year}{2019}).
\newblock \eprint{https://www.science.org/doi/pdf/10.1126/science.aaw4352}.

\bibitem{mamin2013nanoscale}
\bibinfo{author}{Mamin, H.} \emph{et~al.}
\newblock \bibinfo{journal}{\bibinfo{title}{Nanoscale nuclear magnetic resonance with a nitrogen-vacancy spin sensor}}.
\newblock {\emph{\JournalTitle{Science}}} \textbf{\bibinfo{volume}{339}}, \bibinfo{pages}{557--560} (\bibinfo{year}{2013}).

\bibitem{doi:10.1126/science.1231675}
\bibinfo{author}{Staudacher, T.} \emph{et~al.}
\newblock \bibinfo{journal}{\bibinfo{title}{Nuclear magnetic resonance spectroscopy on a (5-nanometer)<sup>3</sup> sample volume}}.
\newblock {\emph{\JournalTitle{Science}}} \textbf{\bibinfo{volume}{339}}, \bibinfo{pages}{561--563}, \doiprefix\url{10.1126/science.1231675} (\bibinfo{year}{2013}).
\newblock \eprint{https://www.science.org/doi/pdf/10.1126/science.1231675}.

\bibitem{doi:10.1126/science.aal2538}
\bibinfo{author}{Lovchinsky, I.} \emph{et~al.}
\newblock \bibinfo{journal}{\bibinfo{title}{Magnetic resonance spectroscopy of an atomically thin material using a single-spin qubit}}.
\newblock {\emph{\JournalTitle{Science}}} \textbf{\bibinfo{volume}{355}}, \bibinfo{pages}{503--507}, \doiprefix\url{10.1126/science.aal2538} (\bibinfo{year}{2017}).
\newblock \eprint{https://www.science.org/doi/pdf/10.1126/science.aal2538}.

\bibitem{doi:10.1126/science.aam8697}
\bibinfo{author}{Aslam, N.} \emph{et~al.}
\newblock \bibinfo{journal}{\bibinfo{title}{Nanoscale nuclear magnetic resonance with chemical resolution}}.
\newblock {\emph{\JournalTitle{Science}}} \textbf{\bibinfo{volume}{357}}, \bibinfo{pages}{67--71}, \doiprefix\url{10.1126/science.aam8697} (\bibinfo{year}{2017}).
\newblock \eprint{https://www.science.org/doi/pdf/10.1126/science.aam8697}.

\bibitem{RevModPhys.89.035002}
\bibinfo{author}{Degen, C.~L.}, \bibinfo{author}{Reinhard, F.} \& \bibinfo{author}{Cappellaro, P.}
\newblock \bibinfo{journal}{\bibinfo{title}{Quantum sensing}}.
\newblock {\emph{\JournalTitle{Rev. Mod. Phys.}}} \textbf{\bibinfo{volume}{89}}, \bibinfo{pages}{035002}, \doiprefix\url{10.1103/RevModPhys.89.035002} (\bibinfo{year}{2017}).

\bibitem{McCullian2020}
\bibinfo{author}{McCullian, B.~A.} \emph{et~al.}
\newblock \bibinfo{journal}{\bibinfo{title}{Broadband multi-magnon relaxometry using a quantum spin sensor for high frequency ferromagnetic dynamics sensing}}.
\newblock {\emph{\JournalTitle{Nature Communications}}} \textbf{\bibinfo{volume}{11}}, \bibinfo{pages}{5229}, \doiprefix\url{10.1038/s41467-020-19121-0} (\bibinfo{year}{2020}).

\bibitem{doi:10.1126/sciadv.abq8158}
\bibinfo{author}{Wang, Z.} \emph{et~al.}
\newblock \bibinfo{journal}{\bibinfo{title}{Picotesla magnetometry of microwave fields with diamond sensors}}.
\newblock {\emph{\JournalTitle{Science Advances}}} \textbf{\bibinfo{volume}{8}}, \bibinfo{pages}{eabq8158}, \doiprefix\url{10.1126/sciadv.abq8158} (\bibinfo{year}{2022}).
\newblock \eprint{https://www.science.org/doi/pdf/10.1126/sciadv.abq8158}.

\bibitem{Meinel2021}
\bibinfo{author}{Meinel, J.} \emph{et~al.}
\newblock \bibinfo{journal}{\bibinfo{title}{Heterodyne sensing of microwaves with a quantum sensor}}.
\newblock {\emph{\JournalTitle{Nature Communications}}} \textbf{\bibinfo{volume}{12}}, \bibinfo{pages}{2737}, \doiprefix\url{10.1038/s41467-021-22714-y} (\bibinfo{year}{2021}).

\bibitem{Taylor2008}
\bibinfo{author}{Taylor, J.~M.} \emph{et~al.}
\newblock \bibinfo{journal}{\bibinfo{title}{High-sensitivity diamond magnetometer with nanoscale resolution}}.
\newblock {\emph{\JournalTitle{Nature Physics}}} \textbf{\bibinfo{volume}{4}}, \bibinfo{pages}{810--816}, \doiprefix\url{10.1038/nphys1075} (\bibinfo{year}{2008}).

\bibitem{PhysRevA.58.2733}
\bibinfo{author}{Viola, L.} \& \bibinfo{author}{Lloyd, S.}
\newblock \bibinfo{journal}{\bibinfo{title}{Dynamical suppression of decoherence in two-state quantum systems}}.
\newblock {\emph{\JournalTitle{Phys. Rev. A}}} \textbf{\bibinfo{volume}{58}}, \bibinfo{pages}{2733--2744}, \doiprefix\url{10.1103/PhysRevA.58.2733} (\bibinfo{year}{1998}).

\bibitem{Kraus2014}
\bibinfo{author}{Kraus, H.} \emph{et~al.}
\newblock \bibinfo{journal}{\bibinfo{title}{Room-temperature quantum microwave emitters based on spin defects in silicon carbide}}.
\newblock {\emph{\JournalTitle{Nature Physics}}} \textbf{\bibinfo{volume}{10}}, \bibinfo{pages}{157--162}, \doiprefix\url{10.1038/nphys2826} (\bibinfo{year}{2014}).

\bibitem{PhysRevApplied.9.024029}
\bibinfo{author}{Wang, X.~S.}, \bibinfo{author}{Zhang, H.~W.} \& \bibinfo{author}{Wang, X.~R.}
\newblock \bibinfo{journal}{\bibinfo{title}{Topological magnonics: A paradigm for spin-wave manipulation and device design}}.
\newblock {\emph{\JournalTitle{Phys. Rev. Appl.}}} \textbf{\bibinfo{volume}{9}}, \bibinfo{pages}{024029}, \doiprefix\url{10.1103/PhysRevApplied.9.024029} (\bibinfo{year}{2018}).

\bibitem{Cenker2021}
\bibinfo{author}{Cenker, J.} \emph{et~al.}
\newblock \bibinfo{journal}{\bibinfo{title}{Direct observation of two-dimensional magnons in atomically thin cri3}}.
\newblock {\emph{\JournalTitle{Nature Physics}}} \textbf{\bibinfo{volume}{17}}, \bibinfo{pages}{20--25}, \doiprefix\url{10.1038/s41567-020-0999-1} (\bibinfo{year}{2021}).

\bibitem{Sun2021}
\bibinfo{author}{Sun, Q.-C.} \emph{et~al.}
\newblock \bibinfo{journal}{\bibinfo{title}{Magnetic domains and domain wall pinning in atomically thin crbr3 revealed by nanoscale imaging}}.
\newblock {\emph{\JournalTitle{Nature Communications}}} \textbf{\bibinfo{volume}{12}}, \bibinfo{pages}{1989}, \doiprefix\url{10.1038/s41467-021-22239-4} (\bibinfo{year}{2021}).

\bibitem{Sun2019}
\bibinfo{author}{Sun, Z.} \emph{et~al.}
\newblock \bibinfo{journal}{\bibinfo{title}{Giant nonreciprocal second-harmonic generation from antiferromagnetic bilayer cri3}}.
\newblock {\emph{\JournalTitle{Nature}}} \textbf{\bibinfo{volume}{572}}, \bibinfo{pages}{497--501}, \doiprefix\url{10.1038/s41586-019-1445-3} (\bibinfo{year}{2019}).

\bibitem{10413860}
\bibinfo{author}{Liu, J.}, \bibinfo{author}{Wu, J.}, \bibinfo{author}{Ren, Z.}, \bibinfo{author}{Yang, S.} \& \bibinfo{author}{Shao, Q.}
\newblock \bibinfo{title}{On-chip zero-field spin wave frequency multiplier and its application on qubit quantum control}.
\newblock In \emph{\bibinfo{booktitle}{2023 International Electron Devices Meeting (IEDM)}}, \bibinfo{pages}{1--4}, \doiprefix\url{10.1109/IEDM45741.2023.10413860} (\bibinfo{year}{2023}).

\end{thebibliography}

\section*{Acknowledgements}

J.Wu, M. Leung, W. Leung, K. Ho, S. Yang were supported by RGC-AOE (AoE/P-701/20), and RGC-GRF (Grant No.16305422). J. Liu, Z. Ren, Q. Shao were supported by National Key R$\&$D Program of China (Grants No.2021YFA1401500), RGC-GRF (Grant No.16303322), and Research Fund of Guangdong-Hong Kong-Macao Joint Laboratory for Intelligent Micro-Nano Optoelectronic Technology (Grant No. 2020B1212030010).

\section*{ Author contributions}
Q. S. and S. Y. conceived the experiment. J. L., Z. R., and J. W. designed the sample. J. L. and Z. R. fabricated the sample. J. W. and J. L. built the measurement setup and performed the measurements with the help of  M. L., W. L., and K. H.. J. W. and S. Y. analyzed the experiment data. X. W., S. Y., and Q. S. developed the theoretical framework. J. L. performed the simulation with the discussion with Z. R. and J. W.. J. W. and S. Y. wrote the article with help from all co-authors.

\section*{ Competing interests}
The authors declare no competing interest.

\end{document}